\documentclass{elsarticle}

\usepackage{amsmath}    

\newcommand{\miniket}[1]{\vert#1\rangle}
\newcommand{\minibra}[1]{\langle#1\vert}

\newcommand{\id}{\;\mbox{$\rm{I} \hspace{-2.0mm}  {\bf I}$}\,}

\begin{document}

\title{Teleportation protocol with non-ideal conditional local operations}

\author[UCC]{C. Di Franco\corref{cor1}}
\ead{cdifranco@caesar.ucc.ie}

\author[QUB]{D. Ballester}

\cortext[cor1]{Corresponding author}

\address[UCC]{Department of Physics, University College Cork, Cork, Republic of Ireland}
\address[QUB]{School of Mathematics and Physics, Queen's University, Belfast BT7 1NN, United Kingdom}

\begin{abstract}
We analyze teleportation protocol when some of receiver's conditional operations are more reliable than others and a non-maximally entangled channel is shared by the two parts. We show that the average fidelity of teleportation can be maximized by choosing properly the basis in which the sender performs her two-qubit measurement.
\end{abstract}


\maketitle

\section{Introduction}

Quantum teleportation has proved to be one of the most striking applications of entanglement as a resource in quantum information processing (QIP)~\cite{teleportation}. By means of a shared entangled channel and a conditional local operation, a receiver (Bob) is able to reconstruct perfectly the unknown state of a qubit given to a sender (Alice), after she performs a joint entangling measurement and communicates classically her result to him. The original idea has been extended along various directions. For instance, the scheme has been generalized to the cases of continuous variables~\cite{cvteleportation} and qudits~\cite{quditsteleportation}. The use of three-particle entanglement in order to transfer the unknown state to either one of two receivers has been studied~\cite{teleportationwithnonmaxtripartite} and a way to implement efficient quantum computation only with linear optics and quantum teleportation has been proposed~\cite{klm}. Considerable effort has been made in demonstrating quantum teleportation experimentally, by means of polarized entangled photons~\cite{zeilingerandboschiteleportation}, squeezed-state entanglement~\cite{polzikteleportation}, liquid-state nuclear magnetic resonance~\cite{laflammeteleportation}, and trapped ions~\cite{blattteleportation}.

In its standard version, the measurement and the conditional operation are considered both ideal. Quantum teleportation in the presence of a non-maximally entangled resource has already been investigated~\cite{darianokimbose,probabilistic}. {\it Here, however, we consider a different scenario, that has not been studied yet: we analyze the case where the conditional operation that Bob needs to perform on his qubit, in order to reconstruct the original state to be teleported, is imperfect}. In particular, we assume that some of the possible operations are more reliable than others. For instance, the identity, which corresponds experimentally to not acting on the state, can be clearly regarded as very accurate. In this Letter, we show that, when the channel is non-maximally entangled, the average fidelity of teleportation can be maximized by a proper selection of the basis used by Alice to perform her measurement. The cases of pure as well as mixed entangled resources are investigated, and different models for decoherence are considered.

This Letter is organized as follows. In Sec.~\ref{protocol}, we describe the protocol for quantum teleportation, in a scenario where the shared channel is pure and non-maximally entangled; {\it the main result of our analysis here is that Alice can choose the basis in which she performs her measurement, in order to cope with Bob's unreliable operations}. In Sec.~\ref{details}, we address explicitly the problem of maximizing teleportation fidelity, and a simple formula for the proper choice of the measurement basis is presented. In Sec.~\ref{mixed} and Sec.~\ref{discussion}, we extend our investigation to cases where the channel is not pure; different models for decoherence are also considered. In this context, we find formulas similar to the ones obtained for pure states. Finally, we summarize our results in Sec.~\ref{remarks}.

\section{General protocol for quantum teleportation with a non-maximally entangled resource}
\label{protocol}

In order to teleport reliably the unknown state of a qubit $T$ from Alice to Bob, an entangled resource has to be shared by them. Let us assume that the state of qubit $T$ is pure (the generalization to a mixed case is, however, straightforward), and equal to
\begin{equation}
 \miniket{\psi}_T=\alpha\,\miniket{0}_T+\beta\,\miniket{1}_T, 
 \label{stateT}
\end{equation}
with $|\alpha|^2+|\beta|^2=1$. For the sake of simplicity, we start by considering that Alice and Bob share a pure non-maximally entangled state (this condition is relaxed in Sec.~\ref{mixed}). A general pure entangled state of two qubits can be written, in an appropriate basis, as
\begin{equation}
 \miniket{\Phi^{(E)}}_{A,B}=\cos\theta\,\miniket{0}_A\miniket{0}_B+\sin\theta\,\miniket{1}_A\miniket{1}_B,
\label{pureentangled}
\end{equation}
where $A$ and $B$ label Alice's and Bob's qubit, respectively, and $\theta \in [0,\pi/4]$ is a parameter measuring the amount of entanglement present in the channel (related to Schmidt coefficients). Alice can perform two-qubit measurements on qubits $T$ (whose initial state has to be teleported) and $A$. The total state of the system can be cast as
\begin{equation}
 \begin{split}
  &\miniket{\psi}_T\miniket{\Phi^{(E)}}_{A,B}=\\
  &(\alpha\,\miniket{0}_T+\beta\,\miniket{1}_T)(\cos\theta\,\miniket{0}_A\miniket{0}_B+\sin\theta\,\miniket{1}_A\miniket{1}_B)=\\
  &\alpha\cos\theta\,\miniket{000}_{T,A,B}+\alpha\sin\theta\,\miniket{011}_{T,A,B}\,+\\
&\beta\cos\theta\,\miniket{100}_{T,A,B}+\beta\sin\theta\,\miniket{111}_{T,A,B}.
 \end{split}
\end{equation}
Let us suppose now that, instead of using the standard basis of Bell states, Alice performs her measurement in the non-maximally entangled orthogonal basis $\{\miniket{\tilde\Phi^\pm},\miniket{\tilde\Psi^\pm}\}$, with
\begin{equation}
 \begin{split}
  &\miniket{\tilde\Phi^+}=\cos\phi\,\miniket{00}+\sin\phi\,\miniket{11},\\
  &\miniket{\tilde\Phi^-}=\sin\phi\,\miniket{00}-\cos\phi\,\miniket{11},\\
  &\miniket{\tilde\Psi^+}=\cos\phi'\,\miniket{01}+\sin\phi'\,\miniket{10},\\
  &\miniket{\tilde\Psi^-}=\sin\phi'\,\miniket{01}-\cos\phi'\,\miniket{10},
 \end{split}
\label{basis}
\end{equation}
and $\phi$, $\phi' \in [0,\pi/2]$. If $\phi=\phi'=\theta$, it is possible to achieve the teleportation of the unknown state $\miniket{\psi}_T$ in Eq.~(\ref{stateT}) with unit fidelity, even though the probability is lower than one whenever $\theta\ne\pi/4$, i.e., the shared resource is non-maximally entangled~\cite{probabilistic}. This is known in literature as probabilistic quantum teleportation. {\it Here, however, we focus our analysis on the maximization of the fidelity for the deterministic scheme}. With the aforementioned basis, the states $\miniket{00}$, $\miniket{01}$, $\miniket{10}$ and $\miniket{11}$ can be written as
\begin{equation}
 \begin{split}
  &\miniket{00}=\cos\phi\,\miniket{\tilde\Phi^+}+\sin\phi\,\miniket{\tilde\Phi^-},\\
  &\miniket{01}=\cos\phi'\,\miniket{\tilde\Psi^+}+\sin\phi'\,\miniket{\tilde\Psi^-},\\
  &\miniket{10}=\sin\phi'\,\miniket{\tilde\Psi^+}-\cos\phi'\,\miniket{\tilde\Psi^-},\\
  &\miniket{11}=\sin\phi\,\miniket{\tilde\Phi^+}-\cos\phi\,\miniket{\tilde\Phi^-}.
 \end{split}
\end{equation}
The total state of the system can thus be recast as
\begin{equation}
 \begin{split}
  &\miniket{\psi}_T\miniket{\Phi^{(E)}}_{A,B}=\\
  &\miniket{\tilde\Phi^+}_{T,A}(\alpha\cos\theta\cos\phi\,\miniket{0}_B+\beta\sin\theta\sin\phi\,\miniket{1}_B)\,+\\
  &\miniket{\tilde\Phi^-}_{T,A}(\alpha\cos\theta\sin\phi\,\miniket{0}_B-\beta\sin\theta\cos\phi\,\miniket{1}_B)\,+\\
  &\miniket{\tilde\Psi^+}_{T,A}(\alpha\sin\theta\cos\phi'\,\miniket{1}_B+\beta\cos\theta\sin\phi'\,\miniket{0}_B)\,+\\
  &\miniket{\tilde\Psi^-}_{T,A}(\alpha\sin\theta\sin\phi'\,\miniket{1}_B-\beta\cos\theta\cos\phi'\,\miniket{0}_B).
 \end{split}
\end{equation}
After the measurement performed by Alice on qubits $T$ and $A$ in the basis $\{\miniket{\tilde\Phi^\pm},\miniket{\tilde\Psi^\pm}\}$, Bob applies a single-qubit operation on $B$, depending on the result obtained by Alice, as shown in the following table:
\begin{equation}
\begin{array}{c c c}
\mbox{Alice's result}&\hspace{1cm}&\mbox{Bob's operation}\vspace{0.25cm} \\ 
\miniket{\tilde\Phi^+}&\hspace{1cm}&\id\\
\miniket{\tilde\Phi^-}&\hspace{1cm}&\sigma_z\\
\miniket{\tilde\Psi^+}&\hspace{1cm}&\sigma_x\\
\miniket{\tilde\Psi^-}&\hspace{1cm}&\sigma_y\\
\end{array}
\nonumber
\end{equation}
For $\phi=\phi'=\theta=\pi/4$, the protocol reduces to the standard one: the state of qubit $B$ after the conditional operation is exactly the same as the initial one of qubit $T$, independently of the result of Alice's measurement. For $\phi=\phi'=\theta\ne\pi/4$, the state $\miniket{\psi}$ is perfectly teleported only when Alice measures qubits $T$ and $A$ in the state $\miniket{\tilde\Phi^-}$ or $\miniket{\tilde\Psi^+}$, as discussed in Ref.~\cite{probabilistic}. 

{\it So far, we have assumed that the conditional operations can be performed ideally by Bob. In what follows, we extend the investigation to cases where this is no longer true}. The formal description of specific errors affecting the output state of $B$ after inaccurate operations is related to the physical implementation of the qubits. In order to obtain an analysis that is independent of the particular experimental setting, let us start describing these errors in the following general and yet interesting way (different models are considered in Sec.~\ref{mixed}): for a generic state, represented by the density matrix $\rho$, the result of an imperfect operation corresponds to the action of decoherence through a depolarizing channel on the output of the ideal operation $\rho_{\rm out}$. In an operator-sum representation, the final state $\rho_{\rm fin}$ after the action of decoherence is thus
\begin{equation}
\rho_{\rm fin}=\sum_{\mu}\hat{K}_{\mu}\rho_{\rm out}\hat{K}_{\mu}^{\dagger}
\end{equation}
with $\{\hat{K}_{\mu}\}$ the set of Kraus operators corresponding to the specific channel, such that $\sum_{\mu}\hat{K}_{\mu}^{\dagger}\hat{K}_{\mu}=\id$~\cite{preskill}. For a depolarizing channel, we have
\begin{equation}
\begin{split}
&\hat{K}_1=\sqrt{1-p}\,\id\;,\;\;\;\hat{K}_2=\sqrt{\frac{p}{3}}\,\sigma_z\;,\\
&\hat{K}_3=\sqrt{\frac{p}{3}}\,\sigma_x\;,\;\;\;\hat{K}_4=\sqrt{\frac{p}{3}}\,\sigma_y.
\end{split}
\label{depokrauss}
\end{equation}
Moreover, we consider that $p$ depends on the conditional operation to be performed. In this way, we take into account the possibility that some operations can be more reliable than others. This can be justified by the fact that different operations require different times to be performed. Even in the case where, instead of applying the conditional operation, one simply rotates the basis of the subsequent operations, this can lead to a different value of $p$ for each rotation. Specifically, the effect of non-ideal operations on qubit $B$ is summarized in the following table~\cite{comment}:
\begin{equation}
\begin{array}{c c l}
\mbox{Ideal operation}&\hspace{1cm}&\mbox{Result of imperfect operation}\vspace{0.25cm}\\ 
\id&\hspace*{1cm}&\rho\rightarrow (1-\frac{4}{3}p_I)\rho+\frac{2}{3}p_I\id\vspace{0.1cm}\\
\sigma_z&\hspace*{1cm}&\rho\rightarrow (1-\frac{4}{3}p_z)\sigma_z\,\rho\,\sigma_z+\frac{2}{3}p_z\id\vspace{0.1cm}\\
\sigma_x&\hspace*{1cm}&\rho\rightarrow (1-\frac{4}{3}p_x)\sigma_x\,\rho\,\sigma_x+\frac{2}{3}p_x\id\vspace{0.1cm}\\
\sigma_y&\hspace*{1cm}&\rho\rightarrow (1-\frac{4}{3}p_y)\sigma_y\,\rho\,\sigma_y+\frac{2}{3}p_y\id\\
\end{array}
\nonumber
\end{equation}

It is known that the effect of fluctuating birefringence in optical fibers can be formally described as a depolarizing channel affecting the polarization of the input light pulse~\cite{banaszek}. Likewise, the model we propose is particularly interesting in the context of QIP with nuclear magnetic resonance (NMR)~\cite{nmr}. Due to the macroscopic nature of the samples, in NMR experiments the standard projective measurements needed for QIP are replaced by ensemble averages~\cite{nmr,bunes}. This means that, instead of pure states, one needs to deal with {\it pseudo-pure states}~\cite{nmr,bunes} of the type
\begin{equation}
\rho_{\delta}=\delta\miniket{\varphi}\minibra{\varphi}+\frac{1-\delta}{2^N}\id
\end{equation}
where $0\leq \delta \leq 1$ and $\miniket{\varphi}$ is an arbitrary pure state in a $2^N$-dimensional space. This state can be interpreted as a statistical mixture in which only a fraction $\delta$ of the qudits is in the pure state $\miniket{\varphi}$ \cite{long}. The actual value of $\delta$ can depend on the experimental technique used. Whereas in thermally distributed samples $\delta$ can be of the order of $10^{-5}$ \cite{bunes}, it can reach values of $0.912$ in experiments on state preparation using parahydrogen and non-thermal techniques \cite{parahy}.

In the remainder of the Letter we show that, for the deterministic protocol with this set of non-ideal operations, the maximal fidelity does not necessarily correspond to the constraint $\phi=\phi'=\theta$. 

The state of qubit $B$ after Bob's conditional operation is not exactly the same as the initial one of qubit $T$. In order to estimate the performance of this teleportation scheme, therefore, we have to evaluate the fidelity as ${\cal F}=\minibra{\psi}\rho_B\miniket{\psi}$, where $\rho_B$ is the final state of qubit $B$ and $\miniket{\psi}=\alpha\,\miniket{0}+\beta\,\miniket{1}$ is the state to teleport. Clearly, the overall fidelity takes into account the different probabilities of the outcomes.

\section{Maximum fidelity with a pure non-maximally entangled resource}
\label{details}
\begin{figure}
\centerline{{\bf (a)}\hskip5.5cm{\bf (b)}}
\vskip0.25cm
\centerline{\includegraphics[width=5.5cm]{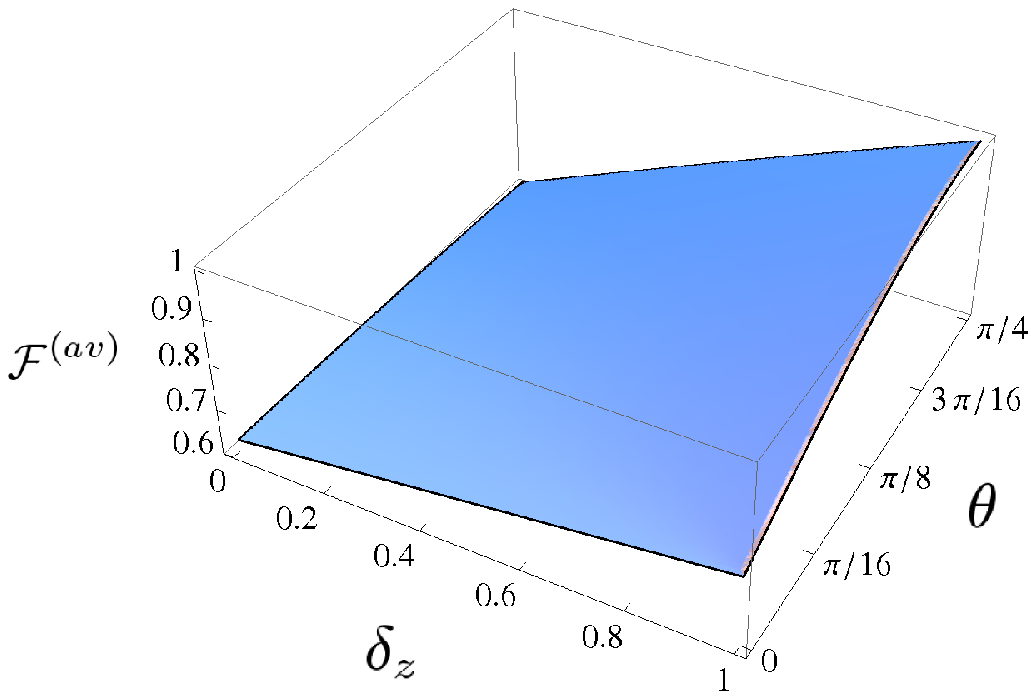}\hskip0.5cm\includegraphics[width=5.5cm]{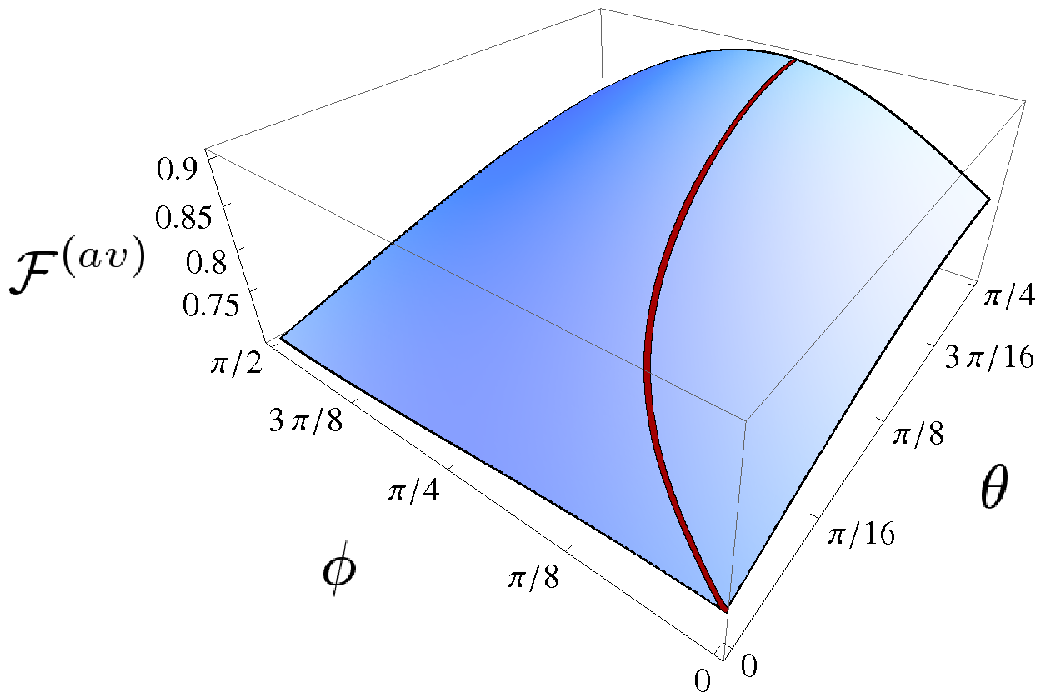}}
\caption{{\bf (a)} Average fidelity of teleportation, for $\delta_I=1$, $\delta_x=\delta_y=\delta_z$, and for the optimal values of $\phi$ and $\phi'$, against the parameters $\delta_z$ and $\theta$. This situation corresponds to the case where the identity operation is ideal and the other three are affected by the same error. {\bf (b)} Average fidelity of teleportation for $\delta_I=1$, $\delta_x=\delta_y=\delta_z=0.9$, and for the optimal value of $\phi'$, against the parameters $\phi$ and $\theta$. The red line corresponds to the optimal value of $\phi$ for a fixed value of $\theta$.}
\label{fav}
\end{figure}

In this Section, the average fidelity of teleportation in the protocol described above is analyzed. It is straightforward to verify that ${\cal F}$ does not depend on any relative phase between the two computational states $\miniket{0}$ and $\miniket{1}$ in the decomposition of $\miniket{\psi}$ in Eq.~(\ref{stateT}). In order to evaluate the average fidelity ${\cal F}^{(av)}$, we can thus parametrize $\alpha$ and $\beta$ as $\cos\gamma$ and $\sin\gamma$, respectively. By assuming a uniform distribution of all the possible input states, we obtain
\begin{equation}
 \begin{split}
  {\cal F}^{(av)}=\,&\frac{1}{2}+\frac{1}{16}\,\{\,\sum_{\nu}\delta_{\nu}+\\
  &\cos2\theta\,[(\delta_I-\delta_z)\cos2\phi+(\delta_x-\delta_y)\cos2\phi'\,]\,+\\
  &\sin2\theta\,[(\delta_I+\delta_z)\sin2\phi+(\delta_x+\delta_y)\sin2\phi'\,]\},
 \end{split}
\label{fidelity}
\end{equation}
with $\delta_{\nu}=1-\frac{4}{3}p_{\nu}$ ($\nu=I,x,y,z$). The behavior of ${\cal F}^{(av)}$, for particular instances of the parameters, is presented in Fig.~\ref{fav}. 

If $p_\nu$ is the same for all the operations, the fidelity is maximized for $\phi=\phi'=\pi/4$, independently of the value of $\theta$. Therefore, Alice needs to perform her measurement in the Bell basis. This is not unexpected, because {\it the effect of changing Alice's basis can be seen as a way to enhance the probability to obtain a specific result with respect to the others}. In this way, she is able to ``privilege" the corresponding conditional operation. If all the operations would give the same error, none of them should be privileged. This is actually what the Bell measurement does, giving the same probability for all the possible results. In the case of maximally entangled channel (i.e., for $\theta=\pi/4$), the average fidelity is also maximized when the measurement is performed in the Bell basis. This can be explained because, when the reduced density matrix of qubit $B$ at the beginning of the protocol corresponds to a totally mixed state (as in the case of a maximally entangled shared resource), there is no way to privilege any operation.

On the other hand, for different values of $p_\nu$ and $\theta\ne\pi/4$, the basis in which Alice should perform her measurement is no longer the Bell one. In general, the maximum average fidelity is obtained for
\begin{equation}
 \begin{split}
  &\tan2\phi=\frac{\delta_I+\delta_z}{\delta_I-\delta_z}\tan2\theta,\\
  &\tan2\phi'=\frac{\delta_x+\delta_y}{\delta_x-\delta_y}\tan2\theta.
 \end{split}
\label{angles}
\end{equation}
One can maximize ${\cal F}^{(av)}$ in Eq.~(\ref{fidelity}) by varying independently the angles $\phi$ and $\phi'$. This is due to the form of the orthogonal basis that has been chosen in Eqs.~(\ref{basis}). This also explains the connection between the pair of operations $\{\id,\sigma_z\}$ and $\{\sigma_x,\sigma_y\}$, which becomes explicit through the optimization relations in Eqs.~(\ref{angles}). As stated above, in the case where $\delta_I=\delta_x=\delta_y=\delta_z$, ${\cal F}^{(av)}$ is optimized for $\phi=\phi'=\pi/4$. For instance, this is what is normally assumed to happen in NMR experiments where weak ensemble measurements are used instead of projective ones. However, if additional depolarization comes into play with different strength depending on the measurement outcome, then this is no longer true.

\section{Maximum fidelity with a mixed non-maximally entangled resource}
\label{mixed}

In order to investigate the described protocol in a way so as to be closer to realistic conditions, we consider now cases where the shared resource is not pure. Let us start with a model that can be easily seen as a generalization of two-qubit Werner states, in which $\miniket{\Phi^{(E)}}$ in Eq.~(\ref{pureentangled}) takes the place of the Bell pair~\cite{werner}. The state of qubits $A$ and $B$ at the beginning of the protocol is then represented by the density matrix
\begin{equation}
 \rho_{AB}=p_W \miniket{\Phi^{(E)}}\minibra{\Phi^{(E)}}+\frac{1-p_W}{4}\id.
 \end{equation}
This shared channel can be, for instance, the result of an initially pure general entangled state (represented by $\miniket{\Phi^{(E)}}$) which is successively affected by white noise, due to the interaction with the environment~\cite{nielsen}. In this case, the average fidelity is
\begin{equation}
 \begin{split}
  {\cal F}^{(av)}=\,&\frac{1}{2}+\frac{1}{16}\,p_W\,\{\,\sum_{\nu}\delta_{\nu}+\\
  &\cos2\theta\,[(\delta_I-\delta_z)\cos2\phi+(\delta_x-\delta_y)\cos2\phi'\,]\,+\\
  &\sin2\theta\,[(\delta_I+\delta_z)\sin2\phi+(\delta_x+\delta_y)\sin2\phi'\,]\}.
 \end{split}
\end{equation}
The only net effect of considering the mixed state $\rho_{AB}$ as the shared resource in the teleportation protocol described in Sec.~\ref{protocol} is a ``global damping" of the fidelity. This implies that the discussion in Sec.~\ref{details} about the optimization of ${\cal F}^{(av)}$, against the angles $\phi$ and $\phi'$, is still valid. In particular, formulas in Eqs.~(\ref{angles}) are applicable also here, i.e., the optimal basis is the same as the previous one.

\begin{figure}
\centerline{{\bf (a)}\hskip5.5cm{\bf (b)}}
\centerline{\includegraphics[width=5.5cm]{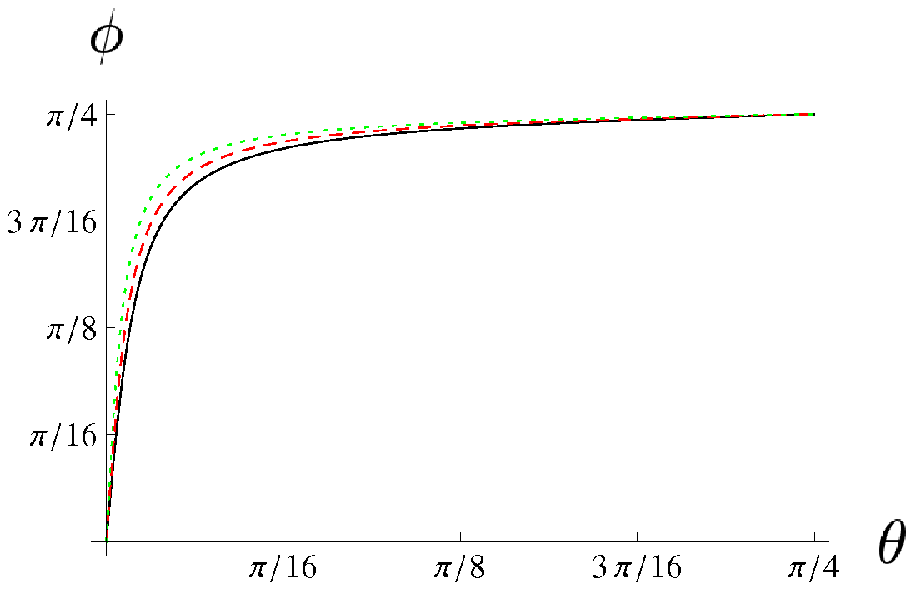}\hskip0.5cm\includegraphics[width=5.5cm]{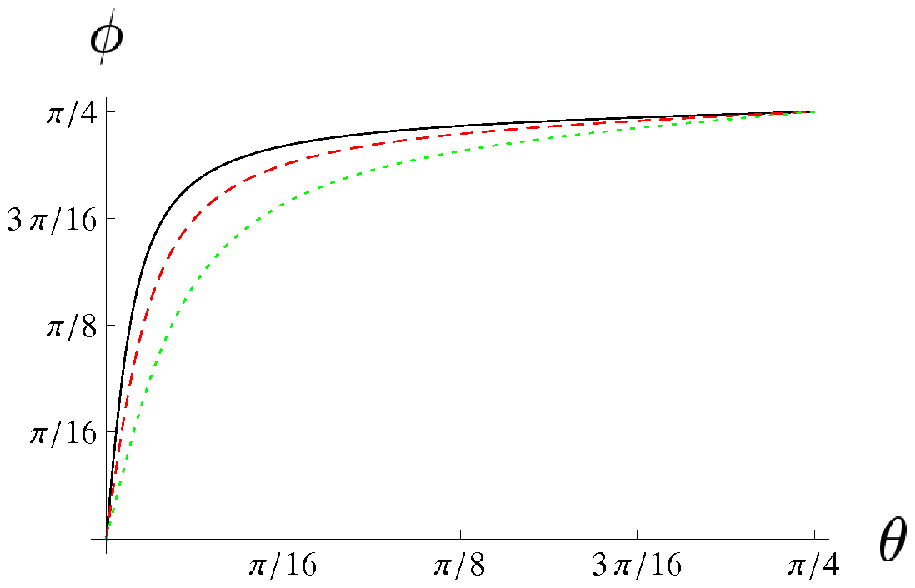}}
\vskip1cm
\centerline{{\bf (c)}}
\centerline{\includegraphics[width=5.5cm]{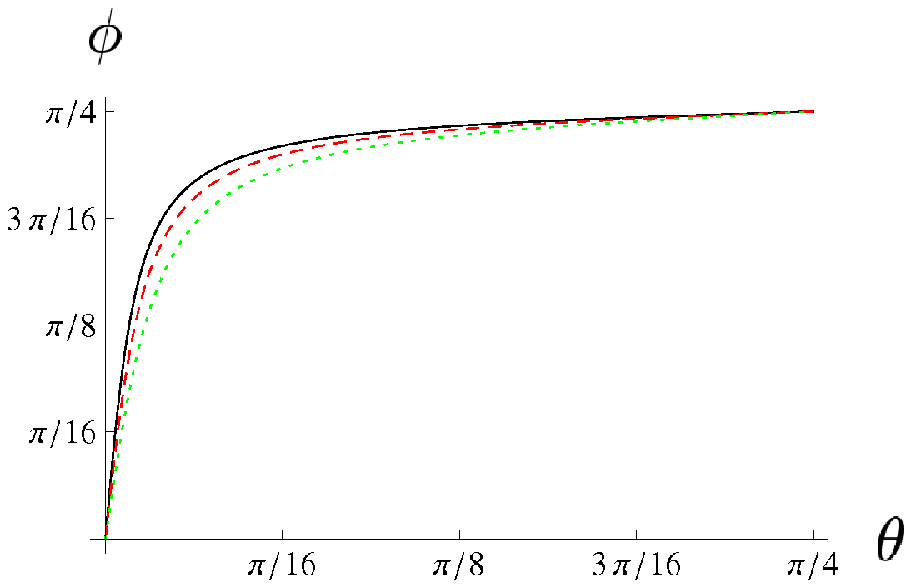}}
\caption{{\bf (a)} Optimal value of $\phi$ against $\theta$, for $\delta_I=1$ and $\delta_Z=0.9$, in the presence of a bit-flip error on the entangled resource. The black, red (dashed) and green (dotted) lines (from the bottom to the top) correspond to $p_{bf}=1$, $p_{bf}=0.9$ and $p_{bf}=0.8$, respectively. {\bf (b)} Optimal value of $\phi$ against $\theta$, for $\delta_I=1$ and $\delta_Z=0.9$, in the presence of a sign-flip error on the entangled resource. The black, red (dashed) and green (dotted) lines (from the top to the bottom) correspond to $p_{sf}=1$, $p_{sf}=0.9$ and $p_{sf}=0.8$, respectively. {\bf (c)} Optimal value of $\phi$ against $\theta$, for $\delta_I=1$ and $\delta_Z=0.9$, in the presence of individual depolarization on the entangled resource. The black, red (dashed) and green (dotted) lines (from the top to the bottom) correspond to $p_{d}=1$, $p_{d}=0.9$ and $p_{d}=0.8$, respectively.}
\label{optangles}
\end{figure}

We can clearly consider different models for decoherence affecting the entangled resource. {\it Due to the fact that the basis $\{\miniket{0},\miniket{1}\}$ for qubits $A$ and $B$ has been chosen in order to write their state as in Eq.~(\ref{pureentangled}) and does not necessarily correspond to a real physical setting (for instance, its components are not the ground and excited state of the single systems), we restrict our analysis to channels affected by computational errors, namely bit- or sign-flips}. Let us suppose first that a bit-flip can occur independently with probability $p_{\rm bf}$ on each qubit of the state $\miniket{\Phi^{(E)}}$~\cite{preskill}. In an operator-sum representation, this can be described by the set of Kraus operators
\begin{equation}
\hat{K}_1=\sqrt{1-p_{\rm bf}}\,\id\;,\;\;\hat{K}_2=\sqrt{p_{\rm bf}}\,\sigma_x.
\end{equation}
For this case we can still evaluate the effect on the average fidelity. With respect to the pure case, however, this is no longer just a global damping. In fact, the terms of ${\cal F}^{(av)}$ in Eq.~(\ref{fidelity}) are now affected in different ways. We have
\begin{equation}
 \begin{split}
  &{\cal F}^{(av)}=\,\frac{1}{2}+\frac{1}{16}\,\{\,(1-2p_{\rm bf})^2\sum_{\nu}\delta_{\nu}+\\
  &(1-2p_{\rm bf})\cos2\theta\,[(\delta_I-\delta_z)\cos2\phi+(\delta_x-\delta_y)\cos2\phi'\,]\,+\\
  &\sin2\theta\,[(\delta_I+\delta_z)\sin2\phi+(\delta_x+\delta_y)\sin2\phi'\,]\}.
 \end{split}
\label{fidelitybf}
\end{equation}
Hence, the optimal angles for Alice's measurement are different from those obtained for the pure shared channel. From Eq.~(\ref{fidelitybf}), one can easily find
\begin{equation}
 \begin{split}
  &\tan2\phi=\frac{1}{1-2p_{\rm bf}}\frac{\delta_I+\delta_z}{\delta_I-\delta_z}\tan2\theta,\\
  &\tan2\phi'=\frac{1}{1-2p_{\rm bf}}\frac{\delta_x+\delta_y}{\delta_x-\delta_y}\tan2\theta.
 \end{split}
\end{equation}
A bit-flip error on the entangled resource has thus the effect to ``push" the optimal angles toward $\pi/4$, corresponding to the Bell measurement, as shown in Fig.~\ref{optangles} {\bf(a)}.

Another error that one can consider is the sign-flip. Also here, we can investigate the case of a sign-flip occurring independently with probability $p_{\rm sf}$ on each qubit. The corresponding set of Kraus operators is
\begin{equation}
\hat{K}_1=\sqrt{1-p_{\rm sf}}\,\id\;,\;\;\hat{K}_2=\sqrt{p_{\rm sf}}\,\sigma_z.
\end{equation}
For the entangled resource that we obtain starting from $\miniket{\Phi^{(E)}}$ in Eq.~(\ref{pureentangled}) and considering the action of a sign-flip error, the formula for the average fidelity of teleportation can be cast as
\begin{equation}
 \begin{split}
  &{\cal F}^{(av)}=\,\frac{1}{2}+\frac{1}{16}\,\{\,\sum_{\nu}\delta_{\nu}+\\
  &\cos2\theta\,[(\delta_I-\delta_z)\cos2\phi+(\delta_x-\delta_y)\cos2\phi'\,]\,+\\
  &(1-2p_{\rm sf})^2\sin2\theta\,[(\delta_I+\delta_z)\sin2\phi+(\delta_x+\delta_y)\sin2\phi'\,]\}.
 \end{split}
\end{equation}
Then, the only effect is the damping factor of the last term in ${\cal F}^{(av)}$. The corresponding angles to maximize the fidelity are
\begin{equation}
 \begin{split}
  &\tan2\phi=(1-2p_{\rm sf})^2\frac{\delta_I+\delta_z}{\delta_I-\delta_z}\tan2\theta,\\
  &\tan2\phi'=(1-2p_{\rm sf})^2\frac{\delta_x+\delta_y}{\delta_x-\delta_y}\tan2\theta.
 \end{split}
\end{equation}
A sign-flip error on the entangled shared resource pushes the optimal angles away from $\pi/4$, as shown in Fig.~\ref{optangles} {\bf(b)}. In this sense, the effect of a sign-flip error is the opposite of the one of a bit-flip error.

Finally, for the sake of completeness, we can also consider depolarization acting individually and independently on the single qubits of the entangled channel, with probability $p_d$. The corresponding Krauss operators are the same as the ones in Eqs.~(\ref{depokrauss}), where $p_d$ takes the role of $p$. For this entangled shared resource, the average fidelity of teleportation is
\begin{equation}
 \begin{split}
  &{\cal F}^{(av)}=\,\frac{1}{2}+\frac{1}{16}\,\{\,(1-2p_d)^2\sum_{\nu}\delta_{\nu}+\\
  &(1-2p_d)\cos2\theta\,[(\delta_I-\delta_z)\cos2\phi+(\delta_x-\delta_y)\cos2\phi'\,]\,+\\
  &(1-2p_d)^2\sin2\theta\,[(\delta_I+\delta_z)\sin2\phi+(\delta_x+\delta_y)\sin2\phi'\,]\}
 \end{split}
\end{equation}
and the angles to maximize it are
\begin{equation}
 \begin{split}
  &\tan2\phi=(1-2p_d)\frac{\delta_I+\delta_z}{\delta_I-\delta_z}\tan2\theta,\\
  &\tan2\phi'=(1-2p_d)\frac{\delta_x+\delta_y}{\delta_x-\delta_y}\tan2\theta.
 \end{split}
\end{equation}
Also for depolarization acting individually on the single qubits of the entangled resource, the optimal measurement moves away from the Bell one, for an increasing value of the decoherence parameter, as shown in Fig.~\ref{optangles} {\bf(c)}.

\section{Different models for Bob's unreliable operations and errors in Alice's measurement}
\label{discussion}
Clearly, the same models for errors can be used to describe also the effect of Bob's unreliable operations. Instead of considering depolarization with different degrees acting on the output of the ideal operations, as previously done in Sec.~\ref{protocol}, one can investigate bit- or sign-flip errors affecting $\rho_{\rm out}$. Interestingly, in both the cases, the formulas for the average fidelity of teleportation and the optimal measurement angles can still be cast in a simple way. Here, for the sake of simplicity, we only report the optimal measurement angles for bit- and sign-flip errors on Bob's output and with the pure non-maximally entangled channel. For bit-flip errors we have
\begin{equation}
 \begin{split}
  &\tan2\phi=\frac{1}{p_{{\rm bf},z}-p_{{\rm bf},I}}\tan2\theta,\\
  &\tan2\phi'=\frac{1}{p_{{\rm bf},y}-p_{{\rm bf},x}}\tan2\theta.
 \end{split}
\label{angles2}
\end{equation}
with $p_{{\rm bf},\nu}$ the probability that a bit-flip occurs on Bob's output state when he performs operation $\nu$ (also in this case, we assume $p_{{\rm bf},I}\le p_{{\rm bf},z}$ and $p_{{\rm bf},x}\le p_{{\rm bf},y}$, see \cite{comment}). On the other hand, when the errors in Bob's operations are described as sign-flips occurring with probabilities $p_{{\rm sf},\nu}$, the optimal measurement always corresponds to the Bell one.

The optimal angles for the cases where the entangled shared resource is mixed, according to the models considered previously in this Letter, and bit-flips occur on Bob's output state can be obtained straightforwardly (for sign-flips, the Bell measurement is always optimal for all the considered models of mixed entangled channels). When the shared resource is affected by bit-flip errors, sign-flip errors or depolarization, one has just to multiply the right-hand side of Eqs.~(\ref{angles2}) by the factors $\frac{1}{1-2p_{\rm bf}}$, $(1-2p_{\rm sf})^2$ and $(1-2p_d)$, respectively.

Let us note that all the formulas derived here for the average fidelity of teleportation can be parametrized in the form
\begin{equation}
\begin{split}
{\cal F}^{(av)}=\,&c_1+c_2\cos2\theta\cos2\phi+c_3\cos2\theta\cos2\phi'+\\
&c_4\sin2\theta\sin2\phi+c_5\sin2\theta\sin2\phi',
\end{split}
\end{equation}
where $c_i$'s depend on the particular model for the entangled resource and Bob's unreliable operations. We can thus also consider a random error in Alice's measurement. Suppose that Alice performs her measurement in a basis with angles $\tilde{\phi}$ and $\tilde{\phi}'$, instead of measuring qubits $T$ and $A$ in the optimal ideal basis according to the rules found so far. Here, we assume that $\tilde{\phi}$ and $\tilde{\phi}'$ are normally distributed around the values $\phi_0$ and $\phi'_0$, with standard deviations $\sigma_{\phi}$ and $\sigma_{\phi'}$, respectively. Then we can average also taking into account all the possible angles in these distributions to obtain 
\begin{equation}
\begin{split}
{\cal F}^{(av)}=\,&c_1+\\
&(e^{-\frac{\sigma_{\phi}^2}{2}}c_2\cos2\theta\cos2\phi_0+e^{-\frac{\sigma_{\phi'}^2}{2}}c_3\cos2\theta\cos2\phi'_0+\\
&e^{-\frac{\sigma_{\phi}^2}{2}}c_4\sin2\theta\sin2\phi_0+e^{-\frac{\sigma_{\phi'}^2}{2}}c_5\sin2\theta\sin2\phi'_0).
\end{split}
\end{equation}
In order to maximize the fidelity, the angles $\phi_0$ and $\phi'_0$ (i.e., the nominal values of the angles that Alice is ``trying" to use in her measurement) have to follow the corresponding rules that we derived throughout the Letter. Thus the results found here are valid also when Alice is not able to perform exact measurements, but she suffers from a random deviation from the ideal basis.

\section{Conclusions}
\label{remarks}

We have considered the effect of different possible imperfections that could arise in the teleportation protocol, such as the use of a non-maximally entangled resource as a channel (pure as well as mixed) and non-ideal operations performed by Bob. {\it The latter, to the best of our knowledge, has never been studied so far}. In particular, in the event of asymmetric errors in these conditional operations, we have found that {\it an optimization of the teleportation fidelity can be performed by properly replacing the Bell basis in the measurement with an orthogonal non-maximally entangled basis}. This result is not only valid for pure entangled resources, but holds also in the case of some mixed shared channels. Similar expressions have been found in the extension of this discussion to different models for the mixed entangled resource and for the description of the unreliable conditional operations, which justifies the interest of our study in a number of experimental scenarios, including the case of random errors in Alice's two-qubit measurement.

\section{Acknowledgments}

We thank G. S. J. Armstrong, M. S. Kim and M. Paternostro for discussions. We acknowledge support from the UK EPSRC and ESF. C.D.F. is supported by the Irish Research Council for Science, Engineering and Technology.


\begin{thebibliography}{99}
\bibitem{comment} Here, we assume $p_I\le p_z$ and $p_x\le p_y$. When these conditions are not satisfied, however, the results presented in this Letter can be still obtained by means of simple sign changes in the definition of the orthogonal basis in Eqs.~(\ref{basis}).
\end{thebibliography}
\end{document}